\documentclass[debug]{rmaa}

\usepackage{paralist}

\usepackage{psfrag,color}

\usepackage[latin1]{inputenc}

\title{Radio Proper Motions of the Nearby Ultra-cool dwarf binary VHS 1256$-$1257AB} 

\author{
  L. F. Rodr\'iguez\altaffilmark{1,2},
  S. A. Dzib\altaffilmark{3},
  L. A. Zapata\altaffilmark{1} 
  L. Loinard\altaffilmark{1} }

\altaffiltext{1}{Instituto de Radioastronom\'\i{}a y Astrof\'\i{}sica,
  UNAM, M\'exico.}

\altaffiltext{2}{Mesoamerican Center for Theoretical Physics, 
UNACH, 
M\'exico}

\altaffiltext{3}{Max-Planck-Institut fur Radioastronomie, Germany}

\shortauthor{Rodr\'iguez, Dzib, Zapata \& Loinard}
\shorttitle{Radio Proper Motions of VHS 1256$-$1257AB}

\fulladdresses{
\item Sergio A. Dzib: Max-Planck-Institut f\"{u}r Radioastronomie, Auf dem H\"{u}gel 69, D-53121 Bonn, Germany. 
\item Laurent Loinard, Luis F. Rodr\'iguez \& Luis A. Zapata: Instituto de Radioastronom\'\i{}a y
  Astrof\'\i{}sica, Universidad Nacional Aut\'onoma de M\'exico,
  Apartado Postal 3--72, 58090 Morelia, Michoac\'an, M\'exico
  (l.rodriguez@irya.unam.mx).}

\listofauthors{L. F. Rodr\'iguez \&
  S. A. Dzib, L. A: Zapata \& L. Loinard}
\indexauthor{Rodr\'iguez, L. F.}
\indexauthor{Dzib, S. A.}
\indexauthor{Zapata, L. A.}
\indexauthor{Loinard, L.}

\abstract{The proper motions of a source obtained at different
epochs or in different spectral regions should
in principle be consistent. However, in the case of a binary source or a source with
associated ejecta, they could be different depending on the epochs
when the observations were made and
on what emission is traced in each spectral region.
In this paper we determine the radio proper motions
of the ultra-cool dwarf binary VHS 1256$-$1257AB from Very Large Array (VLA) observations,
that we find are consistent within error ($\simeq 2-3\%$)
with those reported by Gaia DR3. The comparison of the proper motions and the analysis
of the VLA data imply that, as in the optical, the radio emission is coming in comparable amounts
from both components of the unresolved binary.
}

\resumen{Los movimientos propios de una fuente obtenidos en diferentes \'epocas o en distintas regiones espectrales deben,
en principio, ser consistentes. Sin embargo, en el caso de una fuente binaria o una fuente con eyecciones asociadas,
pueden ser diferentes dependiendo de en que \'epoca se hicieron las observaciones o que emisi\'on es trazada en
cada regi\'on espectral. En este art\'iculo determinamos los movimientos propios en radio de la binaria enana ultrafr\'ia
VHS 1256$-$1257AB a partir de observaciones hechas con el Very Large Array (VLA), encontrando que son dentro
del error  ($\simeq 2-3\%$) consistentes con los valores reportados por Gaia DR3. La comparaci\'on de los movimientos
propios y el an\'alisis de los datos del VLA implican que, como en el \'optico, la emisi\'on de radio proviene en cantidades comparables
de ambos componentes de la binaria no resuelta.}

\addkeyword{Astrometry}
\addkeyword{Radio continuum: stars}
\addkeyword{Stars: binaries: general}
\addkeyword{stars: individual (VHS 1256$-$1257AB)}

\begin{document}
\maketitle

\section{Introduction}
\label{sec:intro}

Highly accurate optical proper motions have been provided by the HIPPARCOS ($\sim 1.2 \times 10^5$ stars;
Perryman et al. 1997; van Leeuwen 2007)
and the Gaia third data release (DR3;  $\sim 1.5 \times 10^9$ stars; Gaia collaboration et al. 2022) 
catalogs. As pointed out by Kervella et al. (2019), discrepancies in the proper 
motions of a source obtained at different epochs (referred to as a proper motion anomaly) could point to 
the presence of a perturbing secondary object. It is of historical relevance to remember that this was the
technique used by Bessel (1844) to infer the existence of a companion (the white dwarf Sirius B) to Sirius.

Besides searching for proper motion anomalies in time at the same spectral
window as done by Kervella et al. (2019), a different approach is to search for them between 
different spectral windows. For example, the
luminosity ratio of the components of a binary system could be very different in different spectral windows. Also
the presence of ejecta with different brightnesses at different frequencies could produce a proper motion anomaly.
An example is the
energetic pulsar PSR J1813$-$1749 for which the X-ray (Chandra) proper motions (Ho et al. 2020) are
much larger than the radio (VLA) proper motions (Dzib \& Rodr{\'\i}guez 2021). Most likely, this 
anomaly is due to the presence of an ejecta that is moving with respect to the host star and is
detectable in the X-rays but not at radio
wavelengths. Then, one can search for proper motion anomalies either between different epochs of time at the
same wavelength or between different spectral windows at the same epoch.

In this paper we present VLA proper motions for the nearby ultra-cool dwarf binary VHS 1256$-$1257AB and compare them with 
the accurate proper motions reported in the Gaia DR3 (Gaia collaboration et al. 2022). 
Our search for proper motion anomalies is restricted to different spectral regions.

\section{VHS 1256$-$1257}
\label{sec:source}

VHS 1256$-$1257AB is a nearby (21.14$\pm$0.22 pc, Gaia collaboration et al. 2022) ultracool dwarf binary composed of
two equal-magnitude stars with a spectral type M7.5$\pm$0.5 (Gauza et al. 2015; Stone et al. 2016). 
This binary system has an orbital period of 7.31$\pm$0.02 yr and a total mass of 0.141$\pm$0.008 $M_\odot$ (Dupuy et al. 2023).
This mass is consistent with the system being a pair of brown dwarfs or of very low-mass stars. Its orbit is highly eccentric ({\it e} = 0.88) with the components reaching a maximum 
separation of $\sim0\rlap.{''}14$
on the plane of the sky (Dupuy et al. 2023). Accurate astrometry of the radio emission could help to better diagnose the spectral and stellar types of the binary dwarfs, although radio emission alone is not a good indicator of spectral type.
It will be more relevant to determine the location and nature of the radio emission. The radio emission could be coming 
from one or both stars and even from the space between them, as has been found in the case of massive binaries 
(i.e. Ortiz-Le{\'o}n et al. 2011). Highly sensitive observations of the spectral index, time variability and polarization 
of the radio emission could discriminate
between thermal (i. e. free-free) or non thermal (i. e. gyrosynchrotron) mechanisms.

In addition,
VHS 1256$-$1257AB hosts a wide-separation planetary-mass companion, VHS 1256$-$1257b, located at $\sim8{''}$ ($\sim$170 au)
to the southwest of the ultracool binary. This companion has a spectral type of L7.0$\pm$1.5 (Gauza et al. 2015) and a 
mass of 19$\pm$5 $M_{Jup}$ (Petrus et al. 2023). VHS 1256$-$1257b is then at the threshold between brown dwarfs and planets
and its potentially planetary mass makes it a target of recent and future 
studies. Indeed, VHS1256-1257b is  the primary spectroscopy target for the Exoplanet Early Release Science program 
of the JWST (Hinkley et al. 2022).

\section{VLA  Observations}
\label{sec:observations}

Ideally, the search for proper motion anomalies should be made by comparing HIPPARCOS with Gaia data.
Unfortunately, VHS 1256$-$1257AB is not included in the much smaller HIPPARCOS catalog. We then determined the proper motions
from less accurate VLA observations. These observations are relevant because this source is not detected in the centimeter with
the European Very-Long-Baseline Interferometry (VLBI) Network,  or in the millimeter
with the NOrthern Extended Millimeter Array (NOEMA) or the Atacama Large Millimetre Array (ALMA; Climent et al. 2022).
We searched in the archives of the 
Karl G. Jansky VLA of NRAO\footnote{The National 
Radio Astronomy Observatory is a facility of the National Science Foundation operated
under cooperative agreement by Associated Universities, Inc.} 
for observation of VHS 1256$-$1257 of good quality and angular resolution made using the same gain calibrator
(in this case J1305$-$1033). This last criterion allows to reach positions that can be compared reliably among different epochs.  

In Table 1 we list the 
three projects found, indicating the epoch of observation, the configuration of the VLA in that epoch, the frequency
and bandwidth observed and the synthesized beam. In this Table we also give the radio flux density 
and position of the unresolved binary VHS 1256$-$1257AB. In no epoch was VHS 1256$-$1257AB
detected in circular polarization. Also, in no epoch was the 
planetary-mass companion VHS 1256$-$1257b detected, either in the I or
V Stokes parameters. These radio data has been analyzed previously by Guirado et al. (2018)
and  Climent et al. (2022). Here we present a combined analysis of the observations that allows a determination of the radio 
proper motions of the source.

\begin{table*}[!t]\centering
  \scriptsize
 \newcommand{\DS}{\hspace{1\tabcolsep}} 
  \begin{changemargin}{-3.0cm}{-2cm}
    \caption{Parameters of the VLA observations of VHS 1256$-$1257AB}
    \setlength{\tabnotewidth}{0.90\linewidth}
    \setlength{\tabcolsep}{1.2\tabcolsep} \tablecols{9}
    \begin{tabular}{l @{\DS} ccccccccc}
      \toprule
      & & VLA & Frequency & Bandwidth & Synthesized & Peak Flux &\multicolumn{2}{c}{Position\tabnotemark{b}}\\
      Project & Epoch & Configuration & (GHz) & (GHz) & Beam & Density\tabnotemark{a} \label{tab:par} & 
      RA(J2000)\tabnotemark{c} \label{tab:par} & DEC(J2000)\tabnotemark{d} \label{tab:par}  \\
      \midrule
      15A-487 & 2015 May 15 &  B$\Rightarrow$BnA  & 10.0 & 3.9 &  
      0$\rlap.{''}$53$\times$0$\rlap.{''}$28; +74$\rlap.^{\circ}$3 & 69.9$ \pm$4.8 & 01$\rlap.^{s}$8478$\pm$0$\rlap.^{s}$0016 & 
      24$\rlap.{''}$8723$\pm$0$\rlap.{''}$0064 \\  
       18A-430 & 2018 Apr 13 &  A  & 6.0 & 3.9 & 0$\rlap.{''}$39$\times$0$\rlap.{''}$22; $-$24$\rlap.^{\circ}$5 & 
       69.6$ \pm$2.6 &
       01$\rlap.^{s}$7947$\pm$0$\rlap.^{s}$0003& 25$\rlap.{''}$4349$\pm$0$\rlap.{''}$0082  \\
       18B-143 & 2018 Nov 17+26 &  C  & 33.1 &  7.8  & 0$\rlap.{''}$72$\times$0$\rlap.{''}$46; +6$\rlap.^{\circ}$4 & 
       65.7$ \pm$9.2 &
       01$\rlap.^{s}$7824$\pm$0$\rlap.^{s}$0022 & 25$\rlap.{''}$5549$\pm$0$\rlap.{''}$0394 \\
           \bottomrule
      \tabnotetext{a}{\scriptsize In $\mu$Jy. } 
       \tabnotetext{b}{\scriptsize Corrected for parallax. }     
       \tabnotetext{c}{\scriptsize Offset from RA(J2000) = $12^h 56^m 00^s$.} 
      \tabnotetext{d}{\scriptsize Offset from DEC(J2000) = $-12^\circ 57' 00''$.}
    \end{tabular}
  \end{changemargin}
\end{table*}

The data were calibrated in the standard manner using the CASA (Common Astronomy Software Applications;  McMullin et al. 2007) package of NRAO and
the pipeline provided for VLA\footnote{https://science.nrao.edu/facilities/vla/data-processing/pipeline} observations. The images 
were made using a robust weighting of 0 (Briggs 1995), searching to optimize the compromise between angular resolution and sensitivity. The positions of VHS 1256$-$1257AB are given in Table 1 and plotted in Figure 1. These positions have
been corrected for the effect of parallax (e.g., Launhardt et al. 2022). The resulting proper motions are 
given in Table 2.

\begin{figure*}[!t]
  \includegraphics[width=0.52\linewidth]{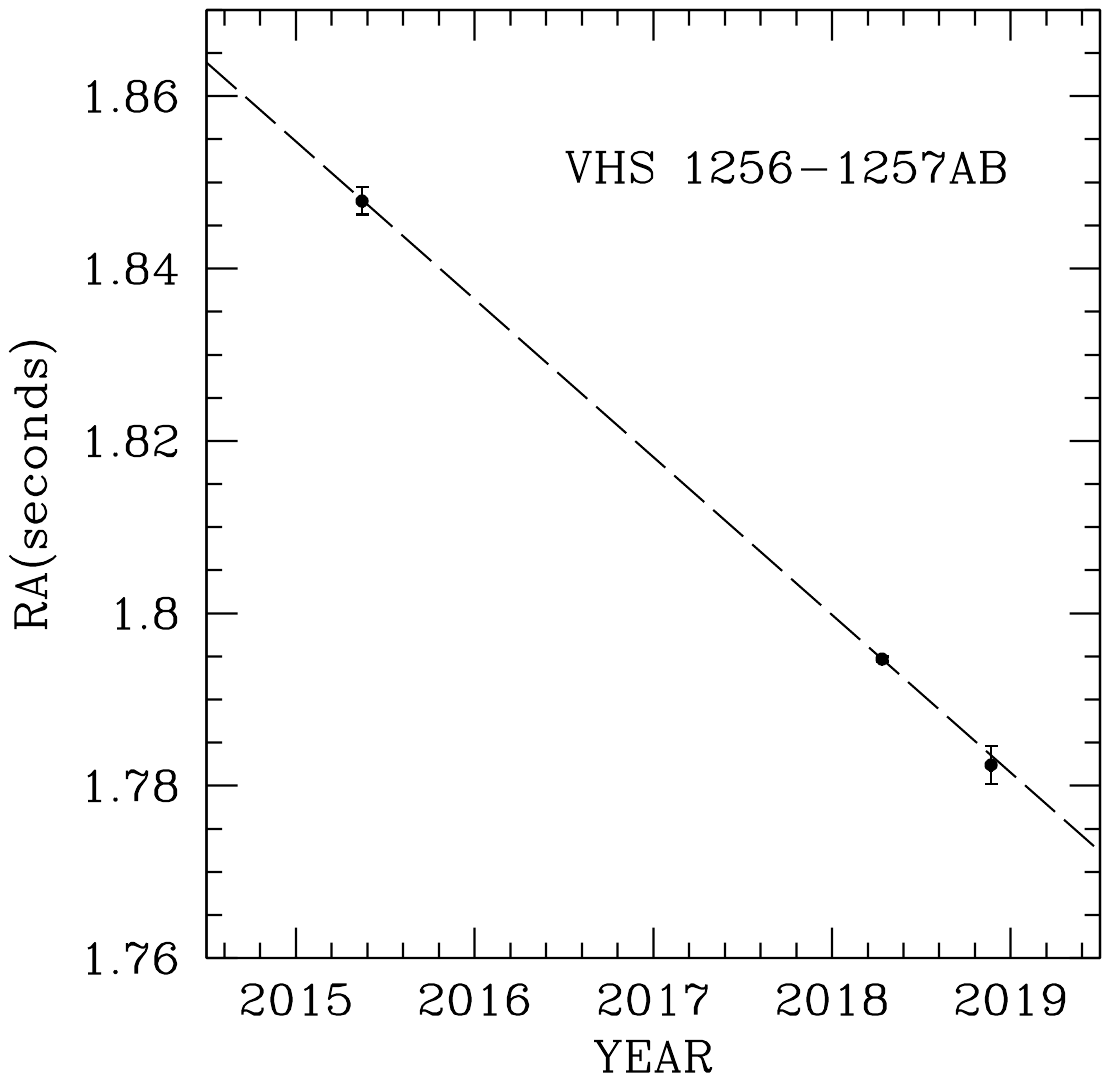}%
  \hfill
  \includegraphics[width=0.52\linewidth]{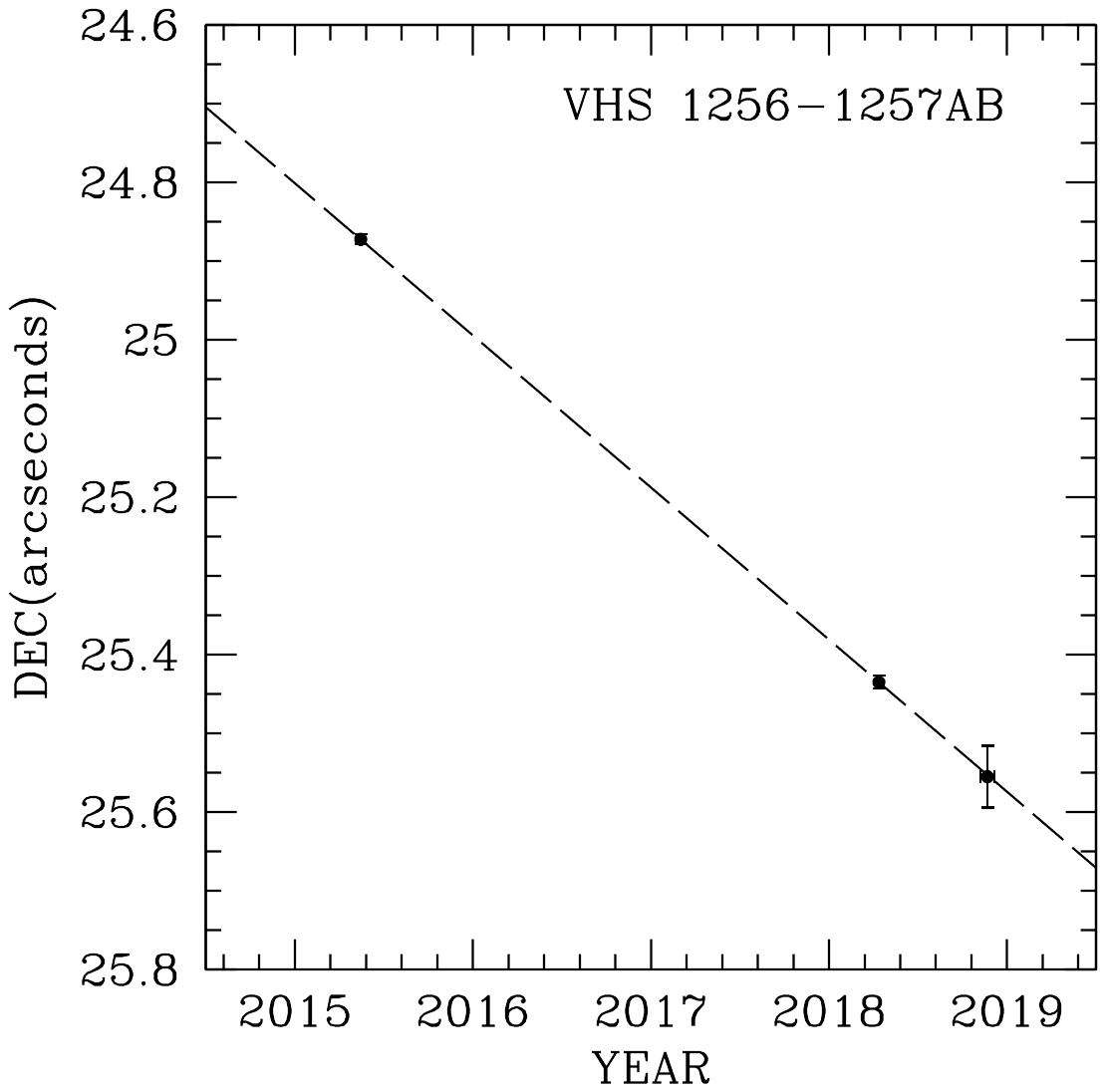}
  \vskip-1.8cm 
  \caption{\scriptsize (Left panel) Right ascension and (right panel) declination of VHS 1256$-$1257AB as a function of time
  for the three epochs analyzed.
  The dashed lines indicate the least squares fit for each parameter. The resulting proper motions are given in Table 2.
  The positions are given as offsets from RA(J2000) = $12^h 56^m 00^s$  and DEC(J2000) = $-12^\circ 57' 00''$}
  \label{fig:pm}
\end{figure*}

\begin{figure*}[!t]
  \includegraphics[width=1.0\linewidth]{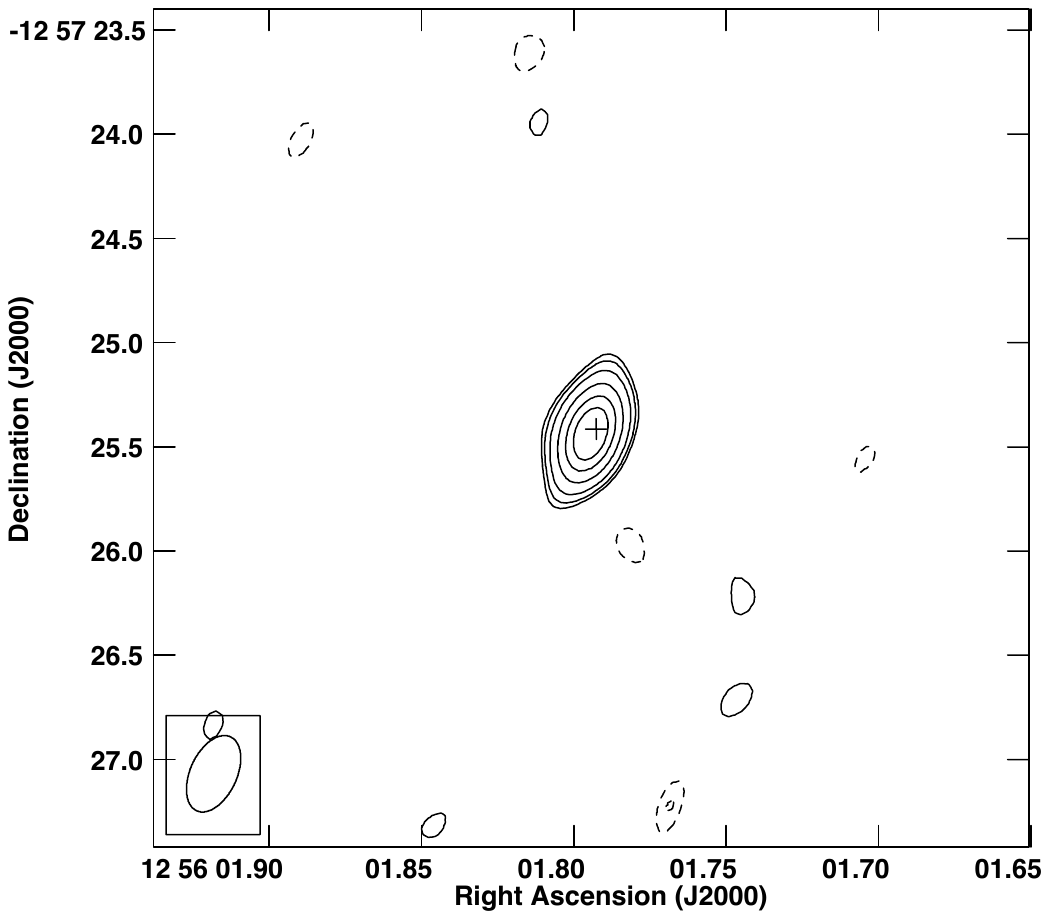}%
  \vskip-2.8cm
    \caption{\scriptsize Very Large Array contour image of VHS 1256$-$1257AB at 6.0 GHz for epoch 2018.28
    corrected for the parallax. Contours 
    are -4, -3, 3, 4, 6, 10, 15 and 20
    times 2.6 $\mu$Jy beam$^{-1}$, the rms noise in this region of the image. The synthesized beam 
    ($0 \rlap.{"}39 \times 0\rlap.{"}22; -24\rlap.^\circ6$)
    is shown in the bottom left corner of the image. The cross marks the Gaia DR3 position of VHS 1256$-$1257AB corrected for
    proper motions and parallax to epoch 2018.28.
    }
  \label{fig:widefig2}
\end{figure*}

\begin{table*}[!t]\centering
  \scriptsize
 \newcommand{\DS}{\hspace{1\tabcolsep}} 
  \begin{changemargin}{-3.0cm}{-2cm}
    \caption{Position and Proper Motions of VHS 1256$-$1257AB}
    \setlength{\tabnotewidth}{0.95\linewidth}
    \setlength{\tabcolsep}{1.2\tabcolsep} \tablecols{6}
    \begin{tabular}{l @{\DS} cccccc}
      \toprule
      & Epoch &\multicolumn{2}{c}{Position\tabnotemark{a} \label{tab:par} } &\multicolumn{2}{c}{Proper Motions\tabnotemark{d} \label{tab:par} }\\
      Telescope & Interval & 
      RA(J2000)\tabnotemark{b} \label{tab:par} & DEC(J2000)\tabnotemark{c} \label{tab:par}  &cos(DEC) $\mu_{RA}$  & $\mu_{DEC}$ \\
      \midrule
      Red Digital Sky Survey & 1956-1997 & 02$\rlap.^{s}$035$\pm$0$\rlap.^{s}$075 & 21$\rlap.{''}$828$\pm$1$\rlap.{''}$113 & -265.0$\pm$47.2 & -195.0$\pm$48.1 \\  
       Gaia DR3 & 2014-2019 & 02$\rlap.^{s}133705$$\pm$0$\rlap.^{s}$000029 & 21$\rlap.{''}92431$$\pm$0$\rlap.{''}$00033 & -272.46$\pm$0.57 &-190.24$\pm$0.50\\
       Very Large Array & 2015-2018 & 02$\rlap.^{s}$1292$\pm$0$\rlap.^{s}$0100& 21$\rlap.{''}$900$\pm$0$\rlap.{''}$058 & -267.6$\pm$8.1 &  -193.4$\pm$3.5\\
           \bottomrule
      \tabnotetext{a}{\scriptsize For epoch 2000.0.}
             \tabnotetext{b}{\scriptsize Offset from RA(J2000) = $12^h 56^m 00^s$.} 
      \tabnotetext{c}{\scriptsize Offset from DEC(J2000) = $-12^\circ 57' 00''$.}
       \tabnotetext{d}{\scriptsize In mas yr$^{-1}$.}
    \end{tabular}
  \end{changemargin}
\end{table*}

\section{Comparison between VLA and Gaia DR3 positions and proper motions}
\label{sec:comparison}
In Table 2 we show the J2000 epoch positions and the proper motions of VHS 1256$-$1257AB from Gaia DR3 
(Gaia collaboration et al. 2022) and the VLA (this paper).
For completeness, we also include these parameters as determined from two images of the Red Digital
Sky Survey taken on 1956 April 07 and 1997 June 02 (Minkowski \& Abell 1968; Djorgovski et al. 1998).
The parameters from the three observations are all consistent at
the $\sim$1-$\sigma$ level and we can rule out the existence of large proper motion anomalies.

Even when the proper motions determined with the VLA have a good precision of $\sim$2-3\%, they are not accurate enough to
apply criteria similar to those used by Kervella et al. (2019) in their comparison of HIPPARCOS and Gaia proper motions.
We then analyze in more detail the optical and radio data with the purpose of comparing the origin of both emissions.

VHS 1256$-$1257AB is formed by two equal-magnitude stars with spectral type M7.5$\pm$0.5 (Gauza et al. 2015; Stone et al. 2016;
Dupuy et al. 2023). We then expect the barycenter and the photocenter of this binary to practically coincide and this explains the 
lack of obvious proper motion anomalies in the optical observations. However, if the radio emission is coming preferentially 
from one of the stars we expect different positions and different proper motions in the optical and in the radio. The fact that the optical and radio proper motions are consistent suggests that, as in the optical, the radio emission is coming from both stars in
comparable amounts.

We now compare the radio and optical positions of VHS 1256$-$1257AB for the epoch 2018 April 13 (=2018.28). We used this 
epoch because it is when the radio data has the highest angular resolution (see Table 1). The radio position is given in Table 1
and the optical Gaia DR3 position after correcting for proper motions and parallax for the same epoch is 
$RA(J2000) = 12^h ~56^m ~01\rlap.^s7927 \pm 00\rlap.^s0007; DEC(J2000) = -12^\circ ~57' ~ 25\rlap.{''}4148 \pm 00\rlap.{''}0094$. 
Most of the error in this position comes from the propagation of the error in the proper motions.
In Figure 2 we show the contour image of the radio emission and
the Gaia DR3 position for this epoch. The differences in the optical and radio positions are 
$\Delta RA(radio - optical) = 0\rlap.^s0020 \pm 0\rlap.^s0008;
~\Delta DEC(radio - optical) = -0\rlap.{''}0201 \pm 0\rlap.{''}0125$.

In epoch 2018.28 the components of VHS 1256$-$1257AB were separated by $\sim0\rlap.{''}13$  at a position angle 
of $\sim156^\circ$; Dupuy et al. 2022). If the radio emission was coming only from one of the components
we expected a difference in the radio and optical declinations of one half of the
separation, $\Delta DEC(radio - optical) = \pm 0\rlap.{''}065$, with the sign of the uncertainty
depending on which of the two stellar components was the radio source. The small difference we found 
($\Delta DEC(radio - optical) = -0\rlap.{''}0201 \pm 0\rlap.{''}0125$) suggests that both stars are emitting radio waves
in similar amounts, perhaps with the 
southern component VHS 1256$-$1257 B being a slightly brighter radio source. This is an interesting result because there is no
\sl a priori \rm reason to expect that stars that are similar in the optical are also similar in the radio.

We now discuss the deconvolved angular size of the 6.0 GHz source associated with VHS 1256$-$1257AB. In the image plane 
we used the CASA task IMFIT to find that the emission can be modeled with a deconvolved Gaussian ellipsoid with dimensions of 
$0\rlap.{''}16 \pm 0\rlap.{''}07 \times 0\rlap.{''}05 \pm 0\rlap.{''}04$  at a position angle of $178^\circ \pm 35^\circ$.
A direct fit to the (u,v) data with the CASA task UVMODELFIT gives a Gaussian ellipsoid with dimensions of 
$0\rlap.{''}15 \pm 0\rlap.{''}02 \times 0\rlap.{''}05 \pm 0\rlap.{''}03$ at a position angle of $178^\circ \pm 9^\circ$. As expected, the fits are consistent
and again support the conclusion that the radio emission is coming from both stellar components of VHS 1256$-$1257AB.
New VLA observations with very high angular resolution and sensitivity will be needed to confirm or refute that the radio emission
is coming from both stars and to investigate the spectral index, variability and polarization of this radio source. Eventually,
the Next Generation VLA will be the ideal instrument for the study of compact radio emission from stars, either single or multiple.

Finally, we note that there is also evidence of binarity from
the RUWE (Renormalized Unit Weight Error) value for VHS 1256$-$1257AB in the Gaia DR3 data. The RUWE is a quality metric provided by the Gaia mission that measures the goodness of fit between the astrometric observations and the astrometric model.
The RUWE is expected to be around 1.0 for sources where the single-star model provides a good fit to the astrometric observations. A value significantly larger than 1.0 could indicate that the source is non-single or otherwise problematic for the astrometric solution. 
For VHS 1256$-$1257AB we have RUWE = 7.3, consistent with the binary nature of the source.

\section{Conclusions}

1) We analyzed VLA observations of the ultra-cool dwarf binary VHS 1256$-$1257AB to obtain its radio proper motions and position.
These parameters are consistent within noise with the ultra-accurate values of Gaia DR3.

2) In combination with the proper motions, the position and angular size of the radio emission from VHS 1256$-$1257AB
are consistent with both stars of the binary emitting comparable radio waves, but with the 
southern component VHS 1256$-$1257 B being somewhat more important. The radio emission alone is not a good
discriminator of spectral type but future sensitive radio observations of the spectral index, time variability and polarization
will help to better understand this dwarf binary.

3) Future very high angular resolution, high sensitivity observations with the VLA can be used to test these conclusions and 
improve our understanding of stellar radio binaries. Eventually,
the Next Generation VLA will be the ideal instrument for the study of compact radio
emission from stars, either single or multiple.

\begin{acknowledgments}


We are thankful to an anonymous referee for valuable comments that improved the manuscript.
This work has made use of data from the European Space Agency (ESA) mission
{\it Gaia} (\url{https://www.cosmos.esa.int/gaia}), processed by the {\it Gaia}
Data Processing and Analysis Consortium (DPAC,
\url{https://www.cosmos.esa.int/web/gaia/dpac/consortium}). Funding for the DPAC
has been provided by national institutions, in particular the institutions
participating in the {\it Gaia} Multilateral Agreement. This work also made use of the 
Digitized Sky Surveys that were produced at the Space Telescope Science Institute under U.S. Government grant NAG W-2166. The images of these surveys are based on photographic data obtained using the Oschin Schmidt Telescope on Palomar Mountain and the UK Schmidt Telescope. L.A.Z. acknowledges financial support from CONACyT-280775 and UNAM-PAPIIT IN110618, 
and IN112323 grants, M\'exico. L.L. acknowledges the support of DGAPA PAPIIT grant IN112820 and CONACYT-CF grant 263356. L.F.R. acknowledges the financial support of DGAPA (UNAM) 
IN105617, IN101418, IN110618 and IN112417 and CONACyT 238631 and 280775-CF grant 263356.

\end{acknowledgments}

\end{document}